\documentclass[sigconf]{acmart}

\AtBeginDocument{%
  \providecommand\BibTeX{{%
    \normalfont B\kern-0.5em{\scshape i\kern-0.25em b}\kern-0.8em\TeX}}}

\begin{document}


\author{Liang Wang, Ying Li, Jierui Zhang, Xianping Tao}
\affiliation{%
  \institution{State Key Laboratory for Novel Software Technology}
  \city{Nanjing University, Nanjing}
  \country{China}
}
\email{wl@nju.edu.cn, mg21070006@smail.nju.edu.cn, jieruizhang@smail.nju.edu.cn, txp@nju.edu.cn}

\renewcommand{\shortauthors}{L. Wang, et al.}

\begin{abstract}
Understanding the evolution of communities in developer social networks (DSNs) around open source software (OSS) projects can provide valuable insights about the socio-technical process of OSS development. Existing studies show the evolutionary behaviors of social communities can effectively be described using patterns including split, shrink, merge, expand, emerge, and extinct. However, existing pattern-based approaches are limited in supporting quantitative analysis, and are potentially problematic for using the patterns in a mutually exclusive manner when describing community evolution. In this work, we propose that different patterns can occur simultaneously between every pair of communities during the evolution, just in different degrees. Four entropy-based indices are devised to measure the degree of community split, shrink, merge, and expand, respectively, which can provide a comprehensive and quantitative measure of community evolution in DSNs. The indices have properties desirable to quantify community evolution including monotonicity, and bounded maximum and minimum values that correspond to meaningful cases. They can also be combined to describe more patterns such as community emerge and extinct. We conduct experiments with real-world OSS projects to evaluate the validity of the proposed indices. The results suggest the proposed indices can effectively capture community evolution, and are consistent with existing approaches in detecting evolution patterns in DSNs with an accuracy of 94.1\%. The results also show that the indices are useful in predicting OSS team productivity with an accuracy of 0.718. In summary, the proposed approach is among the first to quantify the degree of community evolution with respect to different patterns, which is promising in supporting future research and applications about DSNs and OSS development.

\end{abstract}



\title{{Quantitative Analysis of Community Evolution in Developer Social Networks Around Open Source Software Projects}}
\maketitle

\section{Introduction}

Building and maintaining healthy communities around open source software (OSS) projects is crucial because of the socio-technical nature of OSS development \cite{catolino2019gender,ducheneaut2005socialization,weiss2006evolution,nakakoji2002evolution,panichella2014evolution,hannemann2013community,tsay2014influence}.
Cross-disciplinary research on developer social networks (DSNs) that emerge from the developers' communication and cooperation activities in OSS repositories has been conducted by scholars from various fields ever since the early days of the OSS movement \cite{o1999lessons,madey2002open}.
Supported by rich empirical evidences, it is now widely accepted that the social aspects of OSS communities have a strong correlation with, and are important indicators of OSS projects' productivity \cite{wang2007measuring}, success \cite{van2010importance}, and survivability \cite{wang2012survival,ngamkajornwiwat2008exploratory,pinzger2008can}.

While a number of research is about the statistics and patterns of static DSNs snapshots \cite{bird2006mining,crowston2005social,xu2005topological,jensen2011joining,tamburri2019exploring}, there is a series of work that pay special attentions to the evolution of DSNs around OSS projects over time \cite{weiss2006evolution,hannemann2013community,bock2021measuring}.
By understanding the evolution of DSNs, we can not only retrospectively reveal the status of a DSN in a given snapshot, but also predict the future trends of a DSN based on a comprehensive view of its evolution history  \cite{panichella2014evolution,hannemann2013community,bock2021measuring}.
Given the importance of social aspects in OSS development as discussed above, and because OSS development is a ``nature product of evolution'' \cite{nakakoji2002evolution,o1999lessons}, in this work, we study the problem of \emph{measuring the evolution of social communities around OSS projects quantitatively} to support in-depth understanding of the socio-technical aspects of OSS development.

Existing approaches to study community and network evolution can be categorized into two classes which we refer to as \textit{network-level statistics} and \textit{community-level patterns}.
The first class of work uses sequences of statistical metrics over series of temporally-aligned snapshots of social networks to quantify their evolution.
Frequently adopted metrics include network diameter, shortest path between nodes, node betweenness, modularity, hierarchy, centrality, and etc \cite{hannemann2013community,huang2011analysis,aljemabi2018empirical,joblin2017structural,pinzger2008can,schreiber2020social}.
While most of the above metrics measure properties of social networks on a gross level, a natural phenomenon in social networks is that interacting individuals form complex structures organized in \textit{communities}\footnote{For the rest of the paper, a \emph{community} inside a social network refers to a group of interconnected members of the network if not specified otherwise. Refer to Sec. \ref{sec:term} for details about the terms used in this paper.} \cite{palla2007quantifying,bird2008latent}.
Understanding the communities inside DSNs can provide valuable insights about software development.
For instance, the Conway's law suggests that a system's architecture reflects the communication structure of the organization designs it \cite{betz2013evolutionary,conway1968committees}.
Although some metrics such as clustering coefficient \cite{hannemann2013community}, group degree centrality \cite{ngamkajornwiwat2008exploratory}, group betweenness centrality \cite{ngamkajornwiwat2008exploratory}, and etc, are related to measuring substructures or communities \cite{chakraborty2017metrics} in social networks, these network-level statistics are less expressive in revealing the structure and evolution of communities in DSNs.

Different from the above work, the second class of approaches take the community perspective \cite{aljemabi2018empirical} to uncover the evolution patterns of social networks.
Six patterns, including \emph{growth} (\emph{expand}), \emph{contraction} (\emph{shrink}), \emph{merge}, \emph{split}, \emph{birth} (\emph{emerge}), and \emph{death} (\emph{extinct}) are used to describe the evolutionary behaviors of communities inside social networks \cite{hong2011understanding,palla2007quantifying,lin2007blog,asur2009event,brodka2013ged,greene2010tracking,aljemabi2018empirical}.
These patterns are promising in bringing insights about OSS development for corresponding to the evolution of OSS projects and members' activities \cite{hong2011understanding}.
However, the existing pattern-based approaches often assign a single pattern to describe the evolution of a community, which are limited for several reasons.
First, the evolution of a community can hardly be described by a single pattern.
For example, a community can \textit{expand} and \textit{merge} simultaneously with new members and members from other communities joining it.
Choosing either pattern can lead to information loss in the above case.
In addition, matching communities detected in different time steps to identify them as a single, evolving community before pattern assignment is not a trivial task \cite{palla2007quantifying}.
Second, the nominal patterns are promising in describing community evolution qualitatively, but are limited in supporting quantitative analysis.
Third, these patterns are restricted to describe the evolution of a single community. 
It is difficult to extend the concepts to the network level, which is often desirable to gain a comprehensive understanding of the evolution of DSNs.


Our idea to tackle the above limitations can be described as: \emph{the community evolution patterns can happen simultaneously between all pairs of communities in consecutive time steps, only in different degrees.}
Following this idea, we propose in this paper to quantify community evolutionary behaviors comprehensively by calculating \emph{how much} the communities \textit{expand}, \textit{shrink}, \textit{merge}, and \textit{split} simultaneously.
As introduced in Sec. \ref{sec:measure}, the calculation is done by first building a community member migration matrix which captures the dynamics of communities detected in Sec. \ref{sec:dsn_community_detection}.
Based on the matrix, we devise indices based on information entropy to measure the degree of community evolution with respect to the four patterns above.
Community \textit{extinct} and \textit{emerge} are represented by special combinations of the indices.
With this approach, the evolution of a community is described by four indices in each step, which addresses the challenge of finding a unique, most appropriate pattern for community evolution.
It is worth noting that our approach is compatible with existing work \cite{hong2011understanding} for being capable of finding the most appropriate evolution pattern with a simple set of rules as introduced later.
We present the properties of the indices, including the monotonicity, and the upper- and lower-bounds, to show that the indices can correctly reveal complex, real-world community evolutionary behaviors quantitatively.
We also aggregate the community-level indices to describe the overall evolutionary behaviors of communities at the network level, which can support quantitative analysis of community evolution at various scales.


We conduct empirical studies using real-world data collected from thirty-two projects hosted on GitHub to evaluate the validity and usefulness of the proposed community evolution indices.
More specifically, we validate the construct validity of the proposed indices in terms of concurrent and discriminant validity criteria \cite{kanika2015research}.
The results for concurrent validity suggest that the community evolution patterns detected with the proposed indices is consistent with the ones detected following an exiting approach \cite{hong2011understanding}, with an accuracy of 94\%.
By analyzing the properties of the indices, and by calculating the Spearman's Correlation Coefficients of the indices, we find that the indices are capable in measuring different aspects of community evolution.
We further demonstrate the usefulness of the proposed indices by study the correlation between the measured community evolution with OSS project team productivity measured by the count of commits \cite{catolino2019gender,Vasilescu2015Gender}.
The results obtained from regression analysis with linear mixed-effects models suggest that the indices are good predictors of team productivity with an accuracy of 0.718 after log transformation.

The main contributions of this paper are as follows. First, we bridge the gap between nominal, community evolution patterns and quantitative measures with a set of entropy-based indices to support quantitative analysis for community evolution.
Second, we demonstrate versatile ways of using the indices, e.g., aggregating community-level indices to generate network-level quantities, and finding the most appropriate pattern from the indices, which can support various tasks.
Finally, we conduct empirical studies with data from real-world OSS projects to validate the proposed indices, and demonstrate the usefulness of the indices in predicting OSS team productivity.

\begin{figure*}[!t]
  \centering
  \includegraphics[width=0.81\linewidth]{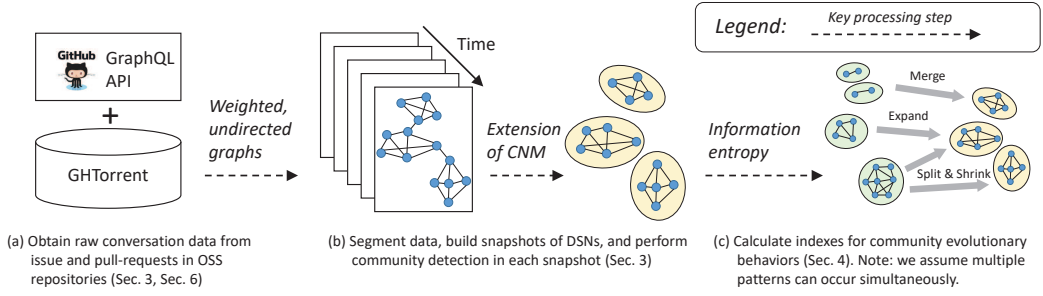}
  \caption{An overview of the proposed approach.}
  \Description{An overview of the proposed approach.}
  \label{fig:overview}
\end{figure*}



\section{Overview of the Proposed Approach}
\label{sec:overview}

This section presents an overview of the proposed approach.

\subsection{Terminology}\label{sec:term}

We first introduce the concepts and terms used in this paper.

An \emph{OSS project} or \emph{repository} refers to a single repository hosted on OSS platforms such as Github, e.g., \emph{tensorflow/tensorflow}\footnote{https://github.com/tensorflow/tensorflow}.
In comparison, large projects, e.g., the Mozilla or GNOME projects that include multiple sub-projects developed together \cite{hong2011understanding,mens2011analysing}, are considered as \emph{ecosystems}  or \emph{super-repositories} following \cite{mens2011analysing,lungu2010small}.
In this work, we focus on community evolution on the \emph{project} level.

A \emph{developer social network} (DSN) around an OSS project include the participants of the project and their relationships, which is typically modeled using a graph model \cite{hong2011understanding,aljemabi2018empirical} extracted from various sources such as mailing lists \cite{bird2006mining,weiss2006evolution,mens2011analysing,panichella2014evolution}, version repositories \cite{mens2011analysing}, issue and bug-tracking systems \cite{panichella2014evolution,aljemabi2018empirical}, and etc.

A \emph{community} inside a DSN means a group of members in the DSN with strongly interconnected links \cite{palla2007quantifying}, or share similar features and actions \cite{nakakoji2002evolution,rossetti2018community}, which is also termed as a \emph{cluster} \cite{hannemann2013community,rossetti2018community}, a \emph{clique} \cite{palla2007quantifying,schreiber2020social}, an \emph{emerging team} \cite{panichella2014evolution}, or a \emph{social group} \cite{palla2007quantifying} in the literature.
We do not use the term to refer to OSS communities in the broad sense \cite{singh2010small}, e.g., the Apache community studied in \cite{weiss2006evolution}.

The \emph{evolutionary behaviors} of communities studied in this paper correspond to the six community evolution patterns studied in \cite{palla2007quantifying,hong2011understanding,lin2007blog,aljemabi2018empirical}, which is different from the general concept of `patterns of community evolution' used in \cite{weiss2006evolution,nakakoji2002evolution}.

\subsection{Overview}

As shown in Fig. \ref{fig:overview}, the processing pipeline proposed in this work involves three steps.
First, as shown in Fig. \ref{fig:overview}(a) and as introduced in Sec. \ref{sec:dsn_community_detection} and Sec. \ref{sec:empirical}, given an OSS repository, we retrieve data about the conversations that take place in issues and pull-requests from the GHTorrent data set \cite{gousios2012ghtorrent} and the GitHub GraphQL API\footnote{https://docs.github.com/en/graphql}.
Participants of the conversations are recorded to extract the joint discussion relationships between them.
The records are used to build DSNs and to perform community evolution analysis in the subsequent steps.
Next, as shown in Fig. \ref{fig:overview}(b) and presented in Sec. \ref{sec:dsn_community_detection},  we segment the records with a siding-window, and build snapshots of DSNs from the segments using weighted, undirected graphs.
Community detection is then performed by applying an extension of the Clauset-Newman-Moore (CNM) algorithm for weighted graphs to the snapshots.
Finally, correspond to Fig. \ref{fig:overview}(c), we measure the evolutionary behaviors, including the \emph{split}, \emph{shrink}, \emph{merge}, and \emph{expand}, of the communities detected in consecutive steps with the information entropy-based indices proposed in Sec. \ref{sec:measure}.
The index values are aggregated to quantify community evolution at the network level before we performing analysis in Sec. \ref{sec:empirical}.


\section{DSN and Community Detection}\label{sec:dsn_community_detection}
In this work, we use a weighted, undirected graph $G= \langle V, E, U, W\rangle$ to model the structure of a DSN, where a vertex $v_i \in V$ represents a user, and an edge $e_{ij} \in E$ represents social interactions between users $v_i$ and $v_j$.
A positive, real-valued weight $u_i \in U$ is assigned to each vertex $v_i$ following Eq. (\ref{eq:node_weight}) to quantify the user's importance. 
And a weight $w_{ij} \in W$ is assigned to each edge $e_{ij}$ following Eq. (\ref{eq:edge_weight})  to quantify the strength of social interactions between the users connected by the edge.
The type of social interaction studied in this paper is the \emph{joint discussion} relationship between users on OSS platforms such as GitHub.
Platform users can post issues and pull-requests to an OSS repository to ask questions and making contributions \cite{bird2008latent,tamburri2019exploring}.
They can also provide responses by leaving comments under posted issues and pull-requests.
We use the term \emph{a conversation} to refer to the collection of the initial body and all the comments to a single issue or pull-request.
There is an edge $e_{ij}$ between two users, $v_i$ and $v_j$, if they join the same conversation.

\subsection{Construct DSN Snapshots}\label{sec:construct_dsn_snapshot}
We construct a series of DSN snapshots from the conversations of an OSS repository.

\subsubsection{Data Segmentation}

Given a series of temporally ordered conversations obtained from a project's repository, $\mathcal{\vec{D}}=\langle d_1, d_2, \cdots, d_N \rangle$, where $d_i=\langle t_i, V_i \rangle, i=1\cdots N$, records the creation time $t_i$, and the set of participants\footnote{The participants of a conversation include the author and the users leave comments in the issue or pull-request.} $V_i$ of the $i$-th record, we apply a non-overlapping sliding window of length $\omega$ to divide $\mathcal{\vec{D}}$ into a series of segments $\mathcal{\vec{S}} = \langle s_1, s_2, \cdots, s_T \rangle$.
The $t$-th segment in $\mathcal{\vec{S}}$ include conversations created in the $t$-th window, i.e., $s_t = \{d_i | t_1 + (t-1), \omega \leq t_i < t_1 + t\omega)\}, j=1\cdots T$, where $t_1$ is the creation time of $d_1$.

\begin{figure*}[!t]
    \centering
    \includegraphics[width = 0.85\linewidth]{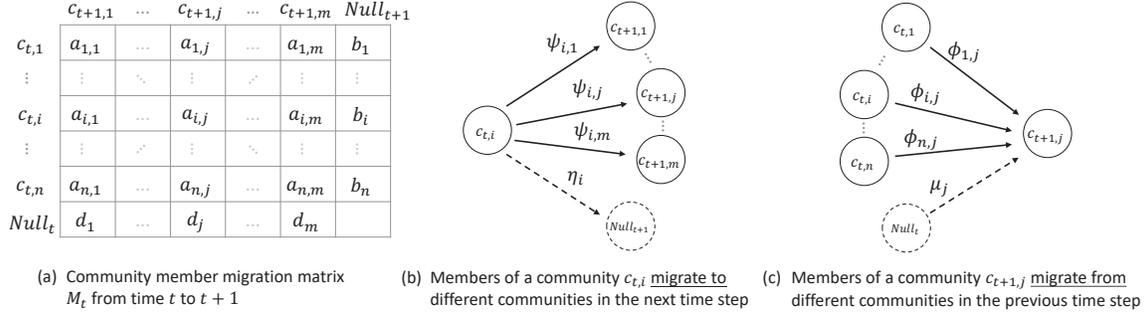}
    \caption{Migration of community members between consecutive time steps. With a slight abuse of notations, we omit the time indicators $t$ and $t+1$ in the matrix shown in (a), and the weights shown in (b) and (c).}
    \label{fig:community_evolution}
\end{figure*}

\subsubsection{Build Local DSNs from Data Segments}\label{sec:local_dsn}

For each segment $s_t\in \mathcal{\vec{S}}$, we build its corresponding local DSN $G^L_t = \langle V^L_t, E^L_t, U^L_t, W^L_t\rangle$ by letting $V^L_t$ to include all the participants of conversations in $s_t$, and $E^L_t$ to include all the joint discussion relationships between the users in $V^L_t$, respectively.
There is an edge between user $v_i$ and $v_j$ if there is at least one conversation $d_k \in s_t$ that both users are participants of $d_k$.
$U^L_t$ and $W^L_t$ include the weights of users and edges in the local DSN, which are calculated as follows.

For each edge $e_{ij}\in E^L_t$, there is a weight $w_{ij}\in W^L_t$ that quantifies the strength of interactions between user $v_i$ and $v_j$.
Let $cnt_{ij}$ be the count of conversations both $v_i$ and $v_j$ participate in segment $s_t$, we assign the weight $w_{ij}$ as follows.
\begin{equation}\label{eq:edge_weight}
    w_{ij} = \ln(cnt_{ij}+1)
\end{equation}
We take the natural logarithm of counts to prevent large counts to overwhelm small ones.
From Eq. (\ref{eq:edge_weight}), we have $w_{ij}=0$ if $cnt_{ij} = 0$, and $w_{ij}>0$ if $cnt_{ij} \geq 0$.
Note we always have $cnt_{ij}=0$ if $i=j$.
Further, we assign a weight $u_i\in U^L_t$ for each user $v_i\in V^L_t$ to quantify the user's importance in the DSN as follows.
\begin{equation}\label{eq:node_weight}
    u_{i} = \ln(\sum_{j=1}^{|V^L_t|} w_{ij} + 1)
\end{equation}
where $|V^L_t|$ is the number of users in segment $s_t$, and $w_{ij}$ is as defined in Eq. (\ref{eq:edge_weight}).
Intuitively, a user with a large weight is a user who actively interactions with many other users, who is considered important in the DSN.
The minimal weight of a user $v_i$ is $u_{i}=0$ if the user does not communicate with any other user in the segment.
With the above approach, we can extract a series of temporally ordered local DSNs $\vec{\mathcal{G}^L} = \langle G^L_1, G^L_2, \cdots, G^L_T \rangle$ from the segments $\mathcal{\vec{S}}$.

\subsubsection{DSN Snapshots with Aggregated Local DSNs}

While it is possible to perform subsequent community detection and evolution analysis with the local DSNs above, it is not trivial to decide the optimal window size $\omega$ for segmentation \cite{hong2011understanding}.
A large window (e.g., from several months to a year \cite{hong2011understanding}) allows us to take a comprehensive snapshot of the DSN with rich data. 
However, it keeps us from performing fine-grained community evolution analysis due to the large time gap between consecutive windows.
On the other hand, a small window allows us to perform fine-grained community evolution analysis.
But a short window may cover too few data to generate a useful snapshot.

To support fine-grained analysis with sufficient data, we generate DSN snapshots by aggregating multiple local DSNs built from one week's data.
We empirically set each snapshot to contain twelve local DSNs to cover data in three months.
We assign more credits to recent data than early data in the snapshot by multiplying a decaying factor to the edge weights of the local DSNs.
Let $G^L_{t-i}$ be the $i^\text{th}$ local DSN being included in the snapshot for time $t$, we decay its edge weights, $W_{t-i}^L$, with a factor $\gamma^i$, where $\gamma$ is heuristically set to $e^{-1}$.
With the above approach, we can obtain a series of DSN snapshots $\vec{\mathcal{G}} = \langle G_1, G_2, \cdots, G_T \rangle$ by aggregating local DSNs.

\subsection{Detect Communities in DSN Snapshots}

The communities studied in this work are groups of users with close and intensive interactions among each other in a snapshot of DSN.
Given the $t$-th snapshot of a DSN, $G_t=\langle V_t, E_t, U_t, W_t \rangle\in\vec{\mathcal{G}}$, we adopt the extension \cite{chakraborty2017metrics,newman2004analysis} of the Clauset-Newman-Moore (CNM) greedy modularity maximization algorithm \cite{newman2003mixing,newman2004finding} to detect the communities within it.
Developed from the traditional CNM algorithm for unweighted graphs, the extended version of CNM performs commnity detection on weighted graphs using the strength of connections between nodes indicated by the edge weights.

The result of community detection for $G_t$ is a set of non-overlapping communities $C_t = \{c_{t,1}, c_{t,2}, \cdots, c_{t,n}\}$, where $c_{t,i} \subseteq V_t$ contains a subset of users in the social network with strong internal joint discussion relationships.
Communities involve only a single member are removed.
By applying community detection to all the snapshots in $\vec{\mathcal{G}}$, we obtain a series of evolving communities over time $\vec{\mathcal{C}} = \langle C_1, C_2, \cdots, C_t, \cdots, C_T\rangle$.
In the next section, we propose metrics to measure community evolution in $\vec{\mathcal{C}}$.


\section{Community Evolution indices}
\label{sec:measure}
Based on the temporally ordered series of communities $\vec{\mathcal{C}}$ detected in the above section, this section presents the process and metrics proposed to measure community evolution.
\subsection{Community Member Migration Matrix}

Given two consecutive sets of communities in $\vec{\mathcal{C}}$, e.g., $C_t$ and $C_{t+1}$, we compute a community member migration matrix to quantify member migration between the communities as shown in Fig. \ref{fig:community_evolution}(a).
Let $n = |C_t|$ and $m = |C_{t+1}|$ be the number of communities in $C_t$ and $C_{t+1}$ (without loss of generality, we assume $n\geq 1$ and $m\geq 1$), respectively, we obtain a $(n+1) \times (m+1)$ matrix $M_t$ where the $ij$-th element $a_{i,j}$ is the number of members migrate from $c_{t,i}$ to $c_{t+1, j}$, calculated following Eq. (\ref{eq:a_ij}), for $i \leq n$ and $j\leq m$.
\begin{equation}\label{eq:a_ij}
    a_{i,j} = c_{t,i} \cap c_{t+1,j}, i=1,2,...,n, j=1,2,...,m
\end{equation}

\noindent where $c_{t,i} \cap c_{t+1,j}$ is the set of common users in $c_{t,i}$ and $c_{t+1,j}$, i.e., the set of users in $c_{t,i}$ who migrate to $c_{t+1,j}$.
Member migration between exiting communities in consecutive time steps correspond to the solid arrows in Fig. \ref{fig:community_evolution}(b) and (c), respectively.


Two $Null$ communities, $Null_{t+1}$ and $Null_t$, are added to $C_{t+1}$ and $C_t$ to include members not involved in these communities.
The $Null_{t+1}$ community involves members of a community in $C_t$ who leave the project in time $t+1$, i.e., members of $C_t$ who do not show in any community in $C_{t+1}$.
And the $Null_{t}$ community include newly joined members in a community in $C_{t+1}$ who do not appear in any community in $C_{t}$.
Elements related to the $Null$ communities are in the last column and row of $M_t$ as illustrated in Fig. \ref{fig:community_evolution}(a), which are calculated following Eq. (\ref{eq:b_ij}).
\begin{equation}\label{eq:b_ij}
\begin{aligned}
&b_{i} = c_{t,i} \setminus \bigcup_{j=1}^m a_{i,j}, 1 \leq i \leq n \\
\text{and }& d_{j} = c_{t+1,j} \setminus \bigcup_{i=1}^n a_{i,j}, 1 \leq j \leq m
\end{aligned}
\end{equation}

\noindent where $\bigcup_{j=1}^m a_{i,j}$ is the set of members in $c_{t, i}$ who stay in the project (and join an arbitrary community) in time $t+1$, and $\bigcup_{i=1}^n a_{i,j}$ is the set of members in $c_{t+1, j}$ who migrate from an existing community in time $t$.
As a result, members from $c_{t,i}$ who leave the project (into $Null_{t+1}$), and members in $c_{t+1,j}$ who are new comers (from $Null_t$) are contained in sets $b_{i} = c_{t,i} \setminus \bigcup_{j=1}^m a_{i,j}$ and $d_{j} = c_{t+1,j} \setminus \bigcup_{i=1}^n a_{i,j}$, which correspond to the dashed arrows in Fig. \ref{fig:community_evolution}(b) and (c), respectively.
We omit time indicators for matrix elements, i.e., $a_{i,j}$, $b_i$, and $d_j$, to keep it concise.

The weights on arrows in Fig. \ref{fig:community_evolution}(b) quantify the distribution of members migrate to different communities from $c_{t,i}$ with respect to their weights at time $t$, which are obtained following Eq. (\ref{eq:psi_eta}).
\begin{equation}
\label{eq:psi_eta} \psi_{i,j} = \frac{f_t(a_{i,j})}{f_t(c_{t,i})}, j=1,2,...,m, \text{ and } \eta_{i} = \frac{f_t(b_{i})}{f_t(c_{t,i})}
\end{equation}
where $a_{i,j}$ and $b_{i}$ are sets of users as defined in Eq. (\ref{eq:a_ij}) and Eq. (\ref{eq:b_ij}), respectively; $c_{t,i}$ is the set of the community's members; let $x$ be a set of users, function $f_t(x)$ is the sum of weights for users in set $x$, and the subscript $t$ indicate the weights are given by snapshot $G_t$:
\begin{equation*}
f_t(x) = \sum_{v_i \in x} u_i, u_i \in U_t, x\subseteq V_t    
\end{equation*}

Similarly, the weights in Fig. \ref{fig:community_evolution}(c) are calculated following Eq. (\ref{eq:phi_mu}) to quantify the distribution of members in $c_{t+1,j}$ from different sources based on their weights in snapshot $G_{t+1}$ given by $f_{t+1}(\cdot)$.
\begin{equation}
\begin{aligned}
\label{eq:phi_mu}  \phi_{i,j} = &\frac{f_{t+1}(a_{i,j})}{f_{t+1}(c_{t+1,j})}, i=1,2,...,n, \text{ and } \mu_{j} = \frac{f_{t+1}(d_{j})}{f_{t+1}(c_{t+1,j})}
\end{aligned}
\end{equation}

\subsection{Entropy-Based Community Evolution Indices for A Single Community} 

In this section, we propose to measure the evolution of a single community between consecutive time steps, i.e., time $t$ and $t+1$, based on the information entropy of the weight distributions shown in Fig. \ref{fig:community_evolution}(b) and (c).
While existing works describe community evolution with six patterns including \emph{split}, \emph{shrink}, \emph{extinct}, \emph{merge}, \emph{expand}, and \emph{emerge} \cite{hong2011understanding}, we omit the \emph{extinct} and \emph{emerge} patterns because they are special cases of community evolution studied in this work, as discussed in Sec. \ref{sec:special_case}.
More importantly, different from existing works \cite{palla2007quantifying,hong2011understanding} that assign a single pattern to each matched pair of communities, we propose that the evolutionary behaviors take place in every pair of communities in consecutive time steps simultaneously, just a matter of degrees.
As a result, we do not require to find the match between prior and post communities.

\subsubsection{The split and shrink indices of community $c_{t,i}$}
We use a community detected in time $t$, i.e., $c_{t,i}$, as shown in Fig. \ref{fig:community_evolution}(b) to explain how the \textit{split} and \textit{shrink} indices are calculated.

Given the $m+1$ communities in the next time step (which include the $m$ detected communities in $C_{t+1}$, and $Null_{t+1}$), and the weight vector $\langle \psi_{i,1}, \psi_{i,2}, \cdots, \psi_{i,m}, \eta_{i} \rangle$, we obtain the distribution of users' weights who stay in the project by normalizing their actual weights:
\begin{equation}\label{eq:normalization_psi}
\hat{\psi}_{i,j} = \frac{\psi_{i,j}}{\sum_{j=1}^m \psi_{i,j}}
\end{equation}
Clearly, we have $\sum_{j=1}^m \hat{\psi}_{i,j} = 1$.
And we can calculate the information entropy \cite{shannon1948mathematical} of the normalized weight vector by:
\begin{equation}
\label{eq:entropy_t} 
\mathcal{H}_{c_{t,i}} = -\sum_{j=1}^m \hat{\psi}_{i,j} \log_2(\hat{\psi}_{i,j})
\end{equation}
With $m$ fixed, the entropy $\mathcal{H}_{c_{t,i}}$ is upper-bounded by the following maximum entropy obtained when $\hat{\psi}_{i,j}=\frac{1}{m}, j=1,2,\cdots,m$:
\begin{equation}\label{eq:max_entropy_t}
\mathcal{H}^*_{t\rightarrow t+1} = -\log_2(\frac{1}{m})
\end{equation}



\noindent Intuitively, $\mathcal{H}_{c_{t,i}}$ measures the degree of $c_{t,i}$'s members spread into different communities (excluding $Null_{t+1}$) in the next step.
And $\mathcal{H}^*_{t\rightarrow t+1} - \mathcal{H}_{c_{t,i}}$ measures the degree of $c_{t,i}$'s members gather to a single community in the next step.
Both $\mathcal{H}_{c_{t,i}}$ and $\mathcal{H}^*_{t\rightarrow t+1} - \mathcal{H}_{c_{t,i}}$ are equal or above zero.
Based on the entropy in Eq. (\ref{eq:entropy_t}), we compute community $c_{t,i}$'s \emph{split} and  \emph{shrink} indices as follows.

The \emph{split index} of $c_{t,i}$, denoted as $\mathcal{I}^{\psi}_{c_{t,i}}$, measures the degree of separation of its members into concrete communities detected in the next step, which is calculated as:

\begin{equation}
\label{eq:split_index} \mathcal{I}^{\psi}_{c_{t,i}} = (1-\eta_{i}) \mathcal{H}_{c_{t,i}}
\end{equation}
where $\eta_{i}$ is defined in Eq. (\ref{eq:psi_eta}).

The \emph{shrink index} of $c_{t,i}$, denoted as $\mathcal{I}^{\eta}_{c_{t,i}}$, measures the degree of its members move to the $Null_{t+1}$ community in the next step:
\begin{equation}
\label{eq:shrink_index} 
\mathcal{I}^{\eta}_{c_{t,i}} = \eta_{i}(\mathcal{H}^*_{t\rightarrow t+1} - \mathcal{I}^{\psi}_{c_{t,i}} + \sigma_{\eta_i})
\end{equation}
\noindent where $\sigma_{\eta_i}$ is defined as follows to guarantee that the shrink index is always positive when $m\geq1$ and $\eta_{i}>0$.
\begin{equation}\label{eq:sigma}
    \sigma_{\eta_i} = 
   \begin{cases}
   0.5 \eta_i &, m = 1,\\0 &, otherwise.
   \end{cases}
  \end{equation}

\subsubsection{Properties of the split and shrink indices}\label{sec:properties}
\begin{figure}[!t]
    \centering
    \includegraphics[width = \linewidth]{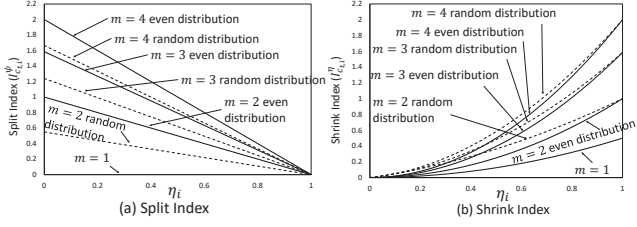}
    \caption{Curves of the \textit{split} and \textit{shrink} indices under different conditions specified by $m$, $\eta_i$, and $\hat{\psi}_{i,j}$. The even distribution corresponds to case $\hat{\psi}_{i,j} = \frac{1}{m},  j=1,\cdots,m$. The random distribution is obtained by randomly assigning  $\hat{\psi}_{i,j}$'s values.
    }
    \label{fig:split_shrink_simulate}
\end{figure}

Fig. \ref{fig:split_shrink_simulate} illustrates the curves of the indices under varying conditions.
The proposed split index, $\mathcal{I}^{\psi}_{c_{t,i}}$, and shrink index, $\mathcal{I}^{\eta}_{c_{t,i}}$, have the following properties that are desirable when measuring community evolution.

$\mathcal{I}^{\psi}_{c_{t,i}}$ and $\mathcal{I}^{\eta}_{c_{t,i}}$ are \underline{strictly monotonic increasing} functions of $m$, given that $\eta_i$ is fixed, and $\hat{\psi}_{i,j} = 1/m, j=1,2,\cdots,m$. Intuitively, the property states that when a fixed portion of members in community $c_{t,i}$ leave the project (i.e., $\eta_i$ fixed), and the remaining members of $c_{t,i}$ who stay in the project migrate evenly to the communities detected in the next step (i.e., $\hat{\psi}_{i,j} = 1/m, j=1,2,\cdots,m$), the split and shrink indices increase with the increasing number of communities detected in the next step (i.e., $m$).

$\mathcal{I}^{\eta}_{c_{t,i}}$ / $\mathcal{I}^{\psi}_{c_{t,i}}$ is a \underline{strictly monotonic increasing / decreasing} function of $\eta_i$, respectively, given that $m>1$ and the distribution of $\hat{\psi}_{i,j}, j=1,2,\cdots,m$ fixed. Intuitively, this property states that the shrink / split index increases / decreases with the increasing portion of users who leave the project (i.e., $\eta_i$), respectively. A special case is that when $m=1$, we have $\mathcal{I}^{\psi}_{c_{t,i}} = 0, \forall \eta_i \in [0,1]$, which is intuitively correct that the community cannot split if there is only one community detected in the next step. For the shrink index $\mathcal{I}^{\eta}_{c_{t,i}}$, we guarantee that $\mathcal{I}^{\eta}_{c_{t,i}}$ is a strictly monotonic increasing function of $\eta_i$ for $m\geq 1$ by introducing the term $\sigma$ as defined in Eq. (\ref{eq:sigma}).
    
Given $m$ and $\eta_i$, the \underline{maximum split index} $\mathcal{I}^{\psi}_{c_{t,i}} = (1-\eta_i)\mathcal{H}^*_{t\rightarrow t+1}$ is obtained when the members of $c_{t,i}$ migrate to the communities detected in the next step with a even distribution, i.e., when we have $\hat{\psi}_{i,j} = 1/m, j=1,2,\cdots,m$.
And the \underline{minimum split index} $\mathcal{I}^{\psi}_{c_{t,i}} = 0$ is obtained when all the members of  $c_{t,i}$ who stay in the project migrate to a single community in the next step, i.e., there exists a $j'$-th community in time $t+1$ that $\hat{\psi}_{i,j'} = 1$ and $\hat{\psi}_{i,j} = 0, \forall j \neq j'$, resulting in $\mathcal{H}_{c_{t,i}}=0$ and $\mathcal{I}^{\psi}_{c_{t,i}} = 0$.
    
Given $m>1$ and $\eta_i$, the \underline{maximum shrink index} $\mathcal{I}^{\eta}_{c_{t,i}} = \eta_i\mathcal{H}^*_{t\rightarrow t+1}$ is obtained when the corresponding split index is minimized, i.e., all stayed members of community $c_{t,i}$ migrate to a single community in the next step.
And the \underline{minimum shrink index} $\mathcal{I}^{\eta}_{c_{t,i}} = \eta_i^2\mathcal{H}^*_{t\rightarrow t+1}$ is obtained when the members of $c_{t,i}$ migrate evenly to communities in time $t+1$. For the special case of $m=1$, the shrink index is only determined by $\eta_i$ and $\sigma_{\eta_i}$.
    


It is straightforward to proof the above properties given the indices' definitions and the properties of information entropy \cite{shannon1948mathematical}.
We omit the proof in this paper due to page limits.

\subsubsection{The merge and expand indices of $c_{t+1,j}$}

Next, to measure community evolution on community $c_{t+1,j}$'s side as shown in Fig. \ref{fig:community_evolution}(c), we develop the \emph{merge} and \emph{expand} indices of $c_{t+1,j}$ following a similar framework as described above.

The \emph{merge index} of $c_{t+1,j}$, denoted as $\mathcal{I}^{\phi}_{c_{t+1,j}}$, measures the degree of its members that come from different source communities in the previous time step, which is calculated as follows.
\begin{equation}
\label{eq:merge_index} \mathcal{I}^{\phi}_{c_{t+1,j}} = (1-\mu_{j}) \mathcal{H}_{c_{t+1,j}}
\end{equation}
where $\mu_j$ and $\mathcal{H}_{c_{t+1,j}}$ are defined in Eq. (\ref{eq:phi_mu}) and Eq. (\ref{eq:entropy_merge_expand}), respectively.

The \emph{expand index} of $c_{t+1,j}$, denoted as $\mathcal{I}^{\mu}_{c_{t+1,j}}$, measures the degree of its members emerge from $Null_{t}$ in the previous step as follows.
\begin{equation}
\label{eq:expand_index} 
\mathcal{I}^{\mu}_{c_{t+1,j}} = \mu_{j}(\mathcal{H}^*_{t+1\leftarrow t} - \mathcal{I}^{\phi}_{c_{t+1,j}} + \sigma_{\mu_j})
\end{equation}
\noindent where $\sigma_{\mu_j}$ is as defined similar to $\sigma_{\eta_i}$ as follows.
\begin{equation*}
    \sigma_{\mu_j} = 
   \begin{cases}
   0.5 \mu_j &, n = 1,\\0 &, otherwise.
   \end{cases}
  \end{equation*}
The entropy values in the above equations are computed by:

\begin{equation}
\label{eq:entropy_merge_expand}
\mathcal{H}^*_{t+1\leftarrow t} = -\log_2(\frac{1}{n}), 
\text{ and }\mathcal{H}_{c_{t+1,j}} = -\sum_{i=1}^n \hat{\phi}_{i,j} \log_2(\hat{\phi}_{i,j})
\end{equation}
where $\hat{\phi}_{i,j} = \frac{\phi_{i,j}}{\sum_{i=1}^n \phi_{i,j}}$ is the normalized weights.
The \textit{merge} and \textit{expand} indices have similar properties as \textit{split} and \textit{shrink}, respectively.
We omit the details to keep the paper concise.

\subsubsection{Community Extinction and Emergence}\label{sec:special_case}
With the above indices, community extinction and emerges are two special combinations of index values as follows: 

The \emph{extinction} of a community $c_{t,i}$ happens if all of its members leave the project and migrate to $Null_{t+1}$, which leads to the case that $\eta_{i}=1$ and $ \mathcal{H}_{c_{t,i}} = 0$. In this case, the \emph{shrink index} takes the maximum value of $\mathcal{I}^{\eta}_{c_{t,i}} = \mathcal{H}^*_{t\rightarrow t+1} + \sigma_{\eta_i}$\footnote{When $\mathcal{I}^{\eta}_{c_{t,i}}$ takes the maximum value, we have $\sigma_{\eta_i} = 0.5$ if $m=1$, and $\sigma_{\eta_i} = 0$ if $m>1$. Similar for the value of  $\sigma_{\mu_j}$ in the case of \textit{emergence}.}. And the \emph{split index} takes the minimum value of $\mathcal{I}^{\psi}_{c_{t,i}} = 0$.

The \emph{emergence} of a new community $c_{t+1,j}$ happens if all its members are from $Null_t$, in which case we have $\mu_{j}=1$, and $\mathcal{H}_{c_{t+1,j}}=0$. In this case, the \emph{expand index} takes the maximum value of $\mathcal{I}^{\mu}_{c_{t+1,j}} = \mathcal{H}^*_{t+1\leftarrow t} + \sigma_{\mu_{j}}$. And the \emph{merge index} takes the minimum value of $\mathcal{I}^{\phi}_{c_{t+1,j}} = 0$.

There are more combinations of the index values that form meaningful cases, which we leave for our future studies.

\subsection{Aggregate Indices for All Communities}

Following the above process, for each community we can obtain indices that quantify its degree of  \textit{split} and \textit{shrink} with respect to the next step, and its degree of \textit{merge} and \textit{expand} with respect to the previous step.
We calculate the aggregated indices for the four patterns across all communities in the snapshot by taking the weighted average of the communities' indices, where the weight for a community is the sum of node weights in the community over the total weight of nodes in all communities.
We can then obtain four aggregated indices that measure the overall, snapshot-level community evolution  for each time step.



\section{Empirical Studies}\label{sec:empirical}

In this section, we conduct empirical studies to evaluate the validity and usefulness of the proposed community evolution measures.

\begin{table*}[!t]
\addtolength{\tabcolsep}{-2pt} 
\centering
\footnotesize
\caption{Overview of the data set, where \# Conv. is the count of conversations, \#Weeks is the time duration of the project's data, and Status indicates whether the project is active by the end of May, 2019.}\label{tbl:data_set}
\begin{tabular}{|clccc|clccc|}
\hline
\textbf{No.} & \textbf{Project} & \textbf{\#Conv.} & \textbf{\#Weeks} & \textbf{Status} & \textbf{No.} & \textbf{Project} & \textbf{\#Conv.} & \textbf{\#Weeks} & \textbf{Status} \\ \hline

1 & angular-ui/bootstrap & 34783 & 342 & inactive & 17 & jekyll/jekyll & 46354 & 530 & active \\ 
2 & apache/couchdb & 9211 & 522 & active & 18 & jessesquires/ JSQMessagesViewController & 10390 & 305 & inactive \\ 
3 & beautify-web/js-beautify & 6151 & 475 & active & 19 & jruby/jruby & 28347 & 527 & active \\ 
4 & boltdb/bolt & 2830 & 221 & inactive & 20 & junit-team/junit4 & 2840 & 501 & active \\ 
5 & carrierwaveuploader/ carrierwave & 10531 & 528 & active & 21 & kevinzhow/PNChart & 939 & 277 & inactive \\ 
6 & celery/celery & 26594 & 521 & active & 22 & mbleigh/acts-as-taggable-on & 3786 & 526 & active \\ 
7 & Compass/compass & 2692 & 484 & inactive & 23 & onevcat/VVDocumenter-Xcode & 854 & 231 & inactive \\ 
8 & davezuko/react-redux-starter-kit & 6138 & 156 & inactive & 24 & prawnpdf/prawn & 5161 & 526 & active \\ 
9 & divio/django-cms & 27416 & 526 & active & 25 & Prinzhorn/skrollr & 3729 & 299 & inactive \\ 
10 & django-extensions/django-extensions & 5716 & 528 & active & 26 & rails/rails & 196487 & 2578 & active \\ 
11 & eczarny/spectacle & 4198 & 456 & inactive & 27 & redis/redis-rb & 4000 & 523 & active \\ 
12 & eventmachine/ eventmachine & 4565 & 526 & active & 28 & sferik/twitter & 3934 & 521 & active \\ 
13 & facebookincubator/create-react-app & 28207 & 79 & inactive & 29 & Shopify/liquid & 6125 & 526 & active \\ 
14 & i3/i3 & 17286 & 532 & active & 30 & sinatra/sinatra & 6598 & 531 & active \\ 
15 & JakeWharton/ ActionBarSherlock & 4860 & 341 & inactive & 31 & sparklemotion/nokogiri & 10516 & 530 & active \\ 
16 & jcjohnson/neural-style & 3328 & 191 & inactive & 32 & thoughtbot/paperclip & 13025 & 524 & active \\ \hline

\end{tabular}
\end{table*}

\subsection{Preliminaries}
\subsubsection{Data Set}

As summarized in Table \ref{tbl:data_set}, we collect a data set involving thirty-two OSS projects on GitHub to conduct the studies.
The projects are randomly selected from a candidate set of popular (with $\#stars\geq 1000$) projects with sufficient data (with $\#Conv.\geq800$)  that cover different topics including programming language (e.g., \textit{rails/rails}), social networks (e.g., \textit{sferik/twitter}), database (e.g., \textit{apache/couchdb}), mobile (e.g., \textit{Prinzhorn/skrollr}), and etc.
Among the projects, twelve inactive projects--determined following the process in \cite{coelho2017modern}--are included to cover projects in different stages.
For each project, we retrieve data starting from its earliest record on the platform to the end of May, 2019 from two sources: the GHTorrent database \cite{gousios2012ghtorrent} and the Github GraphQL API\footnote{https://developer.github.com/v4/}.
The data for issue and pull-request discussions are extracted for community detection and calculating community evolution indices.
We also retrieve the commit history of the repositories to support correlation analysis between community evolution and team productivity.

\subsubsection{Experiment Organization}\label{sec:exp_organization}
We evaluate the construct validity of the proposed indices with the concurrent and discriminant validity criteria following exiting work and guides \cite{raja_defining_2012,meneely_validating_nodate,kanika2015research}.
As defined in \cite{kanika2015research}, construct validity is ``\textit{the extent to which your measure or instrument actually measures what it is theoretically supposed to measure}''.
In this work, we aim at showing the proposed indices are effective in capturing and quantifying community evolutionary behaviors of DSNs around OSS projects.

First, we adopt the concurrent validity criterion to validate the proposed indices.
Concurrent validity accesses the ability of a measure to distinguish between groups it should theoretically be able to distinguish between \cite{raja_defining_2012,kanika2015research}.
In this work, we compare the categorical evolution patterns detected following an existing approach \cite{hong2011understanding} with the patterns detected using the proposed indices.

Next, we evaluate the discriminant validity of the indices which requires the dimensions to be independent or orthogonal to each other.
Based on the properties of the indices discussed in Sec. \ref{sec:properties}, and by further evaluating the discriminant validity of the measures, we aim at showing that the proposed indices are capable of measuring different aspects of community evolution.

Unfortunately, we are unable to assess the convergent validity of the proposed indices because to the best of our knowledge, this is the first work that quantifies the degree of community evolution that include different evolution patterns simultaneously.


Besides construct validity, we also show the usefulness of the proposed indices by studying their correlations with projects' team productivity measured by the count of commits \cite{catolino2019gender,Vasilescu2015Gender}.
Details of the experiments and their results are as follows.



\subsection{Concurrent Validity}\label{sec:concurrent_validity_result}

\begin{table}[!t]
\addtolength{\tabcolsep}{-2pt} 
\centering
\footnotesize
\caption{Rules to determine the evolution patterns of a community, e.g., the $i^\text{th}$ community at time $t$, $c_{t,i}$.}\label{tbl:pattern_detection_rules}
\begin{tabular}{|c|p{0.4\linewidth}|p{0.45\linewidth}|}
\hline
\textbf{Pattern} & \textbf{Existing Approach} \cite{hong2011understanding} & \textbf{Our Approach} \\ \hline
\textit{extinct} & Community $c_{t,i}$ has no matching post community. & $\mathcal{I}^{\psi}_{c_{t,i}} \leq 0.05(\mathcal{H}^*_{t\rightarrow t+1} + \sigma_{\eta_i})$
and
$\mathcal{I}^{\eta}_{c_{t,i}} \geq 0.95(\mathcal{H}^*_{t\rightarrow t+1} + \sigma_{\eta_i})$, i.e., split index close to zero and shrink index close to the maximum value.\\ \hline

\textit{emerge} & Community $c_{t,i}$ has no matching prior community. & 
$\mathcal{I}^{\phi}_{c_{t,i}} \leq 0.05 (\mathcal{H}^*_{t\leftarrow t-1} + \sigma_{\mu_{i}})$
and
$\mathcal{I}^{\mu}_{c_{t,i}} \geq 0.95 (\mathcal{H}^*_{t\leftarrow t-1} + \sigma_{\mu_{i}})$, i.e., merge index close to zero and expand index close to the maximum value.\\ \hline

\textit{split} & Community $c_{t,i}$ has at least two communities in its post community set.&  $\mathcal{I}^{\psi}_{c_{t,i}} > \mathcal{I}^{\eta}_{c_{t,i}}$, i.e., split index greater than shrink index. \\ \hline

\textit{shrink} & Community $c_{t,i}$ has only one community in its post community set, and the size of the post community is less than the size of $c_{t,i}$.$^{\dagger}$ &  $\mathcal{I}^{\psi}_{c_{t,i}} < \mathcal{I}^{\eta}_{c_{t,i}}$, i.e., split index less than shrink index. \\ \hline

\textit{merge} & Community $c_{t,i}$ has at least two communities in its prior community set. & $\mathcal{I}^{\phi}_{c_{t,i}} > \mathcal{I}^{\mu}_{c_{t,i}}$, i.e., merge index greater than expand index. \\ \hline

\textit{expand} & Community $c_{t,i}$ has only one community in its prior community set, and the size of the prior community is less than the size of $c_{t,i}$. & $\mathcal{I}^{\phi}_{c_{t,i}} < \mathcal{I}^{\mu}_{c_{t,i}}$, i.e., merge index less than expand index. \\ \hline

\textit{undefined} & Others. & Others. \\ \hline

\multicolumn{3}{p{\linewidth}}{$^{\dagger}$ The size of a community is the sum of its members' node weights. We also slightly modify the rule for shrink in \cite{hong2011understanding} to compare the size of $c_{t,i}$ with its post community to make the detection results comparable to ours.}

\end{tabular}
\end{table}

\begin{figure*}
    \centering
    \includegraphics[width = \linewidth]{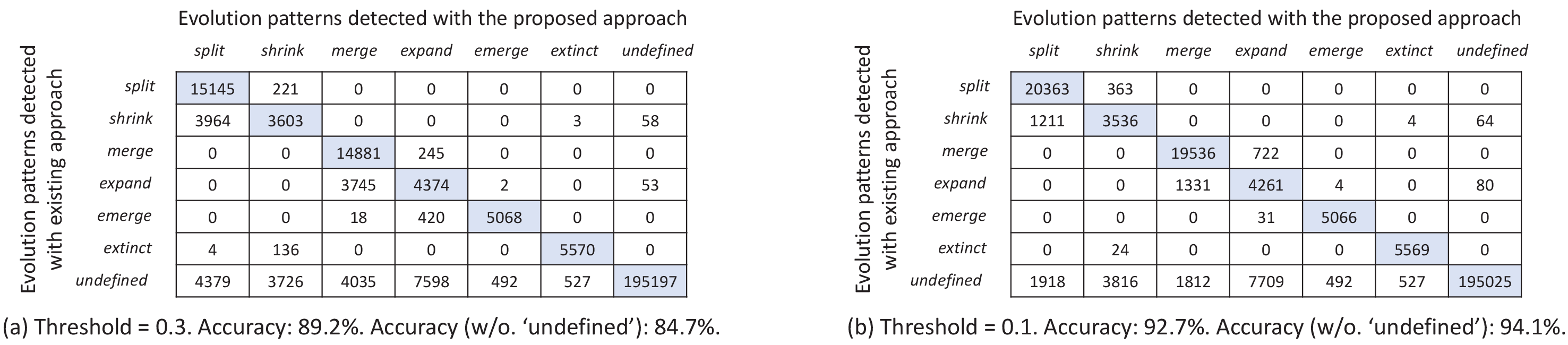}
    \caption{Confusion matrices that compares the community evolution patterns detected with existing approach \cite{hong2011understanding} and the proposed approach following rules specified in Table \ref{tbl:pattern_detection_rules}.}
    \label{fig:confusion_matrix}
\end{figure*}

We evaluate the concurrent validity of the proposed indices by comparing the community evolution pattern detection results of our approach and existing approaches.
We adopt the process proposed in \cite{hong2011understanding} to detect community evolution patterns in DSNs which serve as the `ground-truth' for evaluation.
More specifically, for consecutive time steps $t$ and $t+1$, we first calculate the similarity between each pair of communities detected in the two DSN snapshots.
The generation of DSN snapshots and detection of communities are the same as introduced in Sec. \ref{sec:dsn_community_detection} to conduct a fair comparison.
Slightly different from the original algorithm, the size of a community is defined as the sum of its members' weights instead of the count of its members.
The similarity of two communities is the sum of weights of their common nodes over the smaller value of the two communities' total weights.
We take the threshold of 0.3 proposed in the original work to perform community matching.
Based on the matching results, we adopt rules in Table \ref{tbl:pattern_detection_rules} to determine the evolution patterns for each community.
We modify the rule for determining community \emph{shrink} by comparing a community's size with its matching community in the next step to make the results comparable with ours.
With these rules, for each community, we can assign two evolution patterns to it: 1) one to describe how it evolves towards the next step, which can be \emph{extinct}, \emph{split}, or \emph{shrink}; 2) the other to describe how it evolves from the previous step, which includes \emph{emerge}, \emph{merge}, and \emph{expand}.
An \emph{undefined} pattern is added to label cases that are not included in the existing set of rules.

For community evolution pattern detection using the indices proposed in this paper, we simply compare the index values to determine the patterns of community \emph{split}, \emph{shrink}, \emph{merge}, and \emph{expand} as shown in Table \ref{tbl:pattern_detection_rules}.
For a community, say community $c_{t,i}$, to \emph{extinct}, the ideal case is that its split index is zero ($\mathcal{I}^{\psi}_{c_{t,i}}=0$), and its shrink index takes the maximum value ($\mathcal{I}^{\eta}_{c_{t,i}} = \mathcal{H}^*_{t\rightarrow t+1} + \sigma_{\eta_i}$) as discussed in Sec. \ref{sec:special_case}.
However, this condition is too stringent to effectively detecting community extinction in real-world data.
As a result, we relax the condition as shown in Table \ref{tbl:pattern_detection_rules} that label a community to \emph{extinct} when its split index is close to zero, and its shrink index is close to the maximum value.
The same relaxation also applies to the rule to detect community \emph{emerge}.

The confusion matrix that compares the ground-truth and our detection results is shown in Fig. \ref{fig:confusion_matrix}(a).
The overall accuracy of our approach involving all six patterns and the \emph{undefined} pattern is 89.2\%.
Because the \emph{undefined} pattern is not clearly defined in existing work \cite{hong2011understanding,palla2007quantifying}, we also evaluate the accuracy by removing the \emph{undefined} label (i.e., by removing the last column and row in the confusion matrix).
Without the \emph{undefined} label, the overall accuracy is 84.7\%.
From Fig. \ref{fig:confusion_matrix}(a), we can observe that 52\% of the \emph{shrink} cases are classified as \emph{split} with our approach.
And 45.8\% of the \emph{expand} cases are detected as \emph{merge}.
This is because there is a threshold of 0.3 to filter out matching but less similar communities in the exiting approach.
As a result, the ground-truth is biased to \textit{shrink} and \textit{expand} compared to our approach which does not perform filtering.
Since there is no evidence showing that the threshold of 0.3 is the optimal one, we repeat the test by setting a threshold of 0.1 to increase the chance of community matching when obtaining the ground-truth.
As shown in Fig. \ref{fig:confusion_matrix}(b), we can observe that the detection accuracy with and without the \emph{undefined} pattern increases to 92.7\% and 94.1\%, respectively.

The above results suggest that, with a simple set of rules, the proposed indices can accurately distinguish between different patterns of community evolution which are consistent with the results obtained following the existing approach \cite{hong2011understanding}.
We conclude that the proposed indices for community evolution are valid with respect to the concurrent validity criterion.

\subsection{Discriminant Validity}\label{sec:discriminant_validity_result}

\begin{table}[!t]
\addtolength{\tabcolsep}{-2pt} 
\centering
\footnotesize
\caption{Interpretation of correlation coefficients ($r$).
}\label{tbl:corr_rules}
\begin{tabular}{|c|c|c|}
\hline
\textbf{Levels} & \textbf{Correlation Coefficients ($r$)} & \textbf{Symbols} \\ \hline
Perfect & $r=1$ or $r=-1$  & \colorbox{red}{P}\\ 
Strong & $0.7\leq r < 1$ or $-1<r\leq-0.7$ & \colorbox{pink}{S}\\ 
Moderate & $0.4\leq r < 0.7$ or $-0.7<r\leq-0.4$ & \colorbox{yellow}{M}\\ 
Weak & $0.1\leq r < 0.4$ or $-0.4<r\leq-0.1$ & \colorbox{lime}{W}\\ 
No & $0\leq r < 0.1$ or $-0.1<r\leq 0$ & \colorbox{white}{N}\\ \hline
\end{tabular}
\end{table}

\begin{table}[!t]
\addtolength{\tabcolsep}{-3pt} 
\newcommand{\tabincell}[2]{\begin{tabular}{@{}#1@{}}#2\end{tabular}}
\centering
\tiny
\caption{Pairwise correlation of the aggregated indices (Spearman's Correlation Coefficient). Project No. follow Table \ref{tbl:data_set}. Notations follow Table \ref{tbl:corr_rules}.  The significance levels are shown without the $p$ values to make the table concise.
}\label{tbl:correlation}
\begin{tabular}{|c|llllll|}
\hline
\textbf{Project No.}&\textbf{split-shrink}&\textbf{split-merge}&\textbf{split-expand}&\textbf{shrink-merge}&\textbf{shrink-expand}&\textbf{merge-expand}\\ \hline
1	&	 -0.05 (\colorbox{white}{N})	&	\textcolor{white}{~}0.41 (\colorbox{yellow}{M})***	&	\textcolor{white}{~}0.05 (\colorbox{white}{N})	&	 -0.06 (\colorbox{white}{N})	&	\textcolor{white}{~}0.50 (\colorbox{yellow}{M})***	&	\textcolor{white}{~}0.12 (\colorbox{lime}{W})*\\ 
2	&	\textcolor{white}{~}0.23 (\colorbox{lime}{W})***	&	\textcolor{white}{~}0.73 (\colorbox{pink}{S})***	&	\textcolor{white}{~}0.53 (\colorbox{yellow}{M})***	&	\textcolor{white}{~}0.23 (\colorbox{lime}{W})***	&	\textcolor{white}{~}0.20 (\colorbox{lime}{W})***	&	\textcolor{white}{~}0.54 (\colorbox{yellow}{M})***\\ 
3	&	 -0.07 (\colorbox{white}{N})	&	\textcolor{white}{~}0.20 (\colorbox{lime}{W})***	&	\textcolor{white}{~}0.14 (\colorbox{lime}{W})*	&	 -0.11 (\colorbox{lime}{W})	&	\textcolor{white}{~}0.06 (\colorbox{white}{N})	&	\textcolor{white}{~}0.43 (\colorbox{yellow}{M})***\\ 
4	&	 -0.10 (\colorbox{lime}{W})	&	\textcolor{white}{~}0.39 (\colorbox{lime}{W})***	&	\textcolor{white}{~}0.16 (\colorbox{lime}{W})*	&	 -0.10 (\colorbox{lime}{W})	&	\textcolor{white}{~}0.10 (\colorbox{white}{N})	&	\textcolor{white}{~}0.23 (\colorbox{lime}{W})***\\ 
5	&	\textcolor{white}{~}0.08 (\colorbox{white}{N})	&	\textcolor{white}{~}0.49 (\colorbox{yellow}{M})***	&	\textcolor{white}{~}0.24 (\colorbox{lime}{W})***	&	\textcolor{white}{~}0.06 (\colorbox{white}{N})	&	\textcolor{white}{~}0.26 (\colorbox{lime}{W})***	&	\textcolor{white}{~}0.33 (\colorbox{lime}{W})***\\ 
6	&	\textcolor{white}{~}0.17 (\colorbox{lime}{W})***	&	\textcolor{white}{~}0.59 (\colorbox{yellow}{M})***	&	\textcolor{white}{~}0.25 (\colorbox{lime}{W})***	&	\textcolor{white}{~}0.28 (\colorbox{lime}{W})***	&	\textcolor{white}{~}0.31 (\colorbox{lime}{W})***	&	\textcolor{white}{~}0.26 (\colorbox{lime}{W})***\\ 
7	&	\textcolor{white}{~}0.32 (\colorbox{lime}{W})***	&	\textcolor{white}{~}0.64 (\colorbox{yellow}{M})***	&	\textcolor{white}{~}0.53 (\colorbox{yellow}{M})***	&	\textcolor{white}{~}0.35 (\colorbox{lime}{W})***	&	\textcolor{white}{~}0.39 (\colorbox{lime}{W})***	&	\textcolor{white}{~}0.56 (\colorbox{yellow}{M})***\\ 
8	&	\textcolor{white}{~}0.06 (\colorbox{white}{N})	&	\textcolor{white}{~}0.61 (\colorbox{yellow}{M})***	&	\textcolor{white}{~}0.29 (\colorbox{lime}{W})***	&	\textcolor{white}{~}0.16 (\colorbox{lime}{W})*	&	\textcolor{white}{~}0.12 (\colorbox{lime}{W})	&	\textcolor{white}{~}0.29 (\colorbox{lime}{W})***\\ 
9	&	 -0.07 (\colorbox{white}{N})	&	\textcolor{white}{~}0.52 (\colorbox{yellow}{M})***	&	 -0.04 (\colorbox{white}{N})	&	 -0.01 (\colorbox{white}{N})	&	\textcolor{white}{~}0.10 (\colorbox{white}{N})*	&	\textcolor{white}{~}0.07 (\colorbox{white}{N})\\ 
10	&	\textcolor{white}{~}0.03 (\colorbox{white}{N})	&	\textcolor{white}{~}0.32 (\colorbox{lime}{W})***	&	\textcolor{white}{~}0.19 (\colorbox{lime}{W})***	&	\textcolor{white}{~}0.09 (\colorbox{white}{N})*	&	\textcolor{white}{~}0.15 (\colorbox{lime}{W})***	&	\textcolor{white}{~}0.34 (\colorbox{lime}{W})***\\ 
11	&	\textcolor{white}{~}0.09 (\colorbox{white}{N})	&	\textcolor{white}{~}0.32 (\colorbox{lime}{W})***	&	\textcolor{white}{~}0.41 (\colorbox{yellow}{M})***	&	\textcolor{white}{~}0.07 (\colorbox{white}{N})	&	\textcolor{white}{~}0.15 (\colorbox{lime}{W})**	&	\textcolor{white}{~}0.43 (\colorbox{yellow}{M})***\\ 
12	&	\textcolor{white}{~}0.08 (\colorbox{white}{N})	&	\textcolor{white}{~}0.50 (\colorbox{yellow}{M})***	&	\textcolor{white}{~}0.38 (\colorbox{lime}{W})***	&	\textcolor{white}{~}0.07 (\colorbox{white}{N})	&	\textcolor{white}{~}0.08 (\colorbox{white}{N})	&	\textcolor{white}{~}0.38 (\colorbox{lime}{W})***\\ 
13	&	 -0.20 (\colorbox{lime}{W})	&	\textcolor{white}{~}0.24 (\colorbox{lime}{W})*	&	 -0.22 (\colorbox{lime}{W})	&	\textcolor{white}{~}0.01 (\colorbox{white}{N})	&	\textcolor{white}{~}0.09 (\colorbox{white}{N})	&	 -0.19 (\colorbox{lime}{W})\\ 
14	&	\textcolor{white}{~}0.65 (\colorbox{yellow}{M})***	&	\textcolor{white}{~}0.76 (\colorbox{pink}{S})***	&	\textcolor{white}{~}0.62 (\colorbox{yellow}{M})***	&	\textcolor{white}{~}0.63 (\colorbox{yellow}{M})***	&	\textcolor{white}{~}0.68 (\colorbox{yellow}{M})***	&	\textcolor{white}{~}0.66 (\colorbox{yellow}{M})***\\ 
15	&	\textcolor{white}{~}0.48 (\colorbox{yellow}{M})***	&	\textcolor{white}{~}0.79 (\colorbox{pink}{S})***	&	\textcolor{white}{~}0.68 (\colorbox{yellow}{M})***	&	\textcolor{white}{~}0.42 (\colorbox{yellow}{M})***	&	\textcolor{white}{~}0.39 (\colorbox{lime}{W})***	&	\textcolor{white}{~}0.66 (\colorbox{yellow}{M})***\\ 
16	&	\textcolor{white}{~}0.11 (\colorbox{lime}{W})	&	\textcolor{white}{~}0.66 (\colorbox{yellow}{M})***	&	\textcolor{white}{~}0.33 (\colorbox{lime}{W})***	&	\textcolor{white}{~}0.15 (\colorbox{lime}{W})*	&	\textcolor{white}{~}0.24 (\colorbox{lime}{W})***	&	\textcolor{white}{~}0.43 (\colorbox{yellow}{M})***\\ 
17	&	\textcolor{white}{~}0.03 (\colorbox{white}{N})	&	\textcolor{white}{~}0.64 (\colorbox{yellow}{M})***	&	\textcolor{white}{~}0.10 (\colorbox{lime}{W})*	&	\textcolor{white}{~}0.12 (\colorbox{lime}{W})**	&	\textcolor{white}{~}0.13 (\colorbox{lime}{W})**	&	\textcolor{white}{~}0.10 (\colorbox{lime}{W})*\\ 
18	&	\textcolor{white}{~}0.32 (\colorbox{lime}{W})***	&	\textcolor{white}{~}0.65 (\colorbox{yellow}{M})***	&	\textcolor{white}{~}0.52 (\colorbox{yellow}{M})***	&	\textcolor{white}{~}0.40 (\colorbox{yellow}{M})***	&	\textcolor{white}{~}0.42 (\colorbox{yellow}{M})***	&	\textcolor{white}{~}0.51 (\colorbox{yellow}{M})***\\ 
19	&	\textcolor{white}{~}0.11 (\colorbox{lime}{W})*	&	\textcolor{white}{~}0.67 (\colorbox{yellow}{M})***	&	\textcolor{white}{~}0.16 (\colorbox{lime}{W})***	&	\textcolor{white}{~}0.14 (\colorbox{lime}{W})**	&	\textcolor{white}{~}0.17 (\colorbox{lime}{W})***	&	\textcolor{white}{~}0.25 (\colorbox{lime}{W})***\\ 
20	&	\textcolor{white}{~}0.35 (\colorbox{lime}{W})***	&	\textcolor{white}{~}0.79 (\colorbox{pink}{S})***	&	\textcolor{white}{~}0.63 (\colorbox{yellow}{M})***	&	\textcolor{white}{~}0.34 (\colorbox{lime}{W})***	&	\textcolor{white}{~}0.34 (\colorbox{lime}{W})***	&	\textcolor{white}{~}0.58 (\colorbox{yellow}{M})***\\ 
21	&	\textcolor{white}{~}0.07 (\colorbox{white}{N})	&	\textcolor{white}{~}0.50 (\colorbox{yellow}{M})***	&	\textcolor{white}{~}0.34 (\colorbox{lime}{W})***	&	\textcolor{white}{~}0.06 (\colorbox{white}{N})	&	\textcolor{white}{~}0.09 (\colorbox{white}{N})	&	\textcolor{white}{~}0.37 (\colorbox{lime}{W})***\\ 
22	&	\textcolor{white}{~}0.04 (\colorbox{white}{N})	&	\textcolor{white}{~}0.43 (\colorbox{yellow}{M})***	&	\textcolor{white}{~}0.26 (\colorbox{lime}{W})***	&	\textcolor{white}{~}0.11 (\colorbox{lime}{W})**	&	\textcolor{white}{~}0.12 (\colorbox{lime}{W})**	&	\textcolor{white}{~}0.31 (\colorbox{lime}{W})***\\ 
23	&	\textcolor{white}{~}0.27 (\colorbox{lime}{W})***	&	\textcolor{white}{~}0.54 (\colorbox{yellow}{M})***	&	\textcolor{white}{~}0.45 (\colorbox{yellow}{M})***	&	\textcolor{white}{~}0.14 (\colorbox{lime}{W})*	&	\textcolor{white}{~}0.16 (\colorbox{lime}{W})*	&	\textcolor{white}{~}0.37 (\colorbox{lime}{W})***\\ 
24	&	\textcolor{white}{~}0.02 (\colorbox{white}{N})	&	\textcolor{white}{~}0.48 (\colorbox{yellow}{M})***	&	\textcolor{white}{~}0.30 (\colorbox{lime}{W})***	&	 -0.03 (\colorbox{white}{N})	&	\textcolor{white}{~}0.07 (\colorbox{white}{N})	&	\textcolor{white}{~}0.33 (\colorbox{lime}{W})***\\ 
25	&	\textcolor{white}{~}0.22 (\colorbox{lime}{W})***	&	\textcolor{white}{~}0.25 (\colorbox{lime}{W})***	&	\textcolor{white}{~}0.36 (\colorbox{lime}{W})***	&	\textcolor{white}{~}0.27 (\colorbox{lime}{W})***	&	\textcolor{white}{~}0.33 (\colorbox{lime}{W})***	&	\textcolor{white}{~}0.50 (\colorbox{yellow}{M})***\\ 
26	&	\textcolor{white}{~}0.16 (\colorbox{lime}{W})***	&	\textcolor{white}{~}0.41 (\colorbox{yellow}{M})***	&	\textcolor{white}{~}0.23 (\colorbox{lime}{W})***	&	\textcolor{white}{~}0.25 (\colorbox{lime}{W})***	&	\textcolor{white}{~}0.24 (\colorbox{lime}{W})***	&	\textcolor{white}{~}0.24 (\colorbox{lime}{W})***\\ 
27	&	 -0.00 (\colorbox{white}{N})	&	\textcolor{white}{~}0.56 (\colorbox{yellow}{M})***	&	\textcolor{white}{~}0.43 (\colorbox{yellow}{M})***	&	 -0.01 (\colorbox{white}{N})	&	\textcolor{white}{~}0.10 (\colorbox{white}{N})*	&	\textcolor{white}{~}0.38 (\colorbox{lime}{W})***\\ 
28	&	 -0.03 (\colorbox{white}{N})	&	\textcolor{white}{~}0.44 (\colorbox{yellow}{M})***	&	\textcolor{white}{~}0.32 (\colorbox{lime}{W})***	&	\textcolor{white}{~}0.03 (\colorbox{white}{N})	&	\textcolor{white}{~}0.14 (\colorbox{lime}{W})**	&	\textcolor{white}{~}0.34 (\colorbox{lime}{W})***\\ 
29	&	\textcolor{white}{~}0.28 (\colorbox{lime}{W})***	&	\textcolor{white}{~}0.75 (\colorbox{pink}{S})***	&	\textcolor{white}{~}0.48 (\colorbox{yellow}{M})***	&	\textcolor{white}{~}0.26 (\colorbox{lime}{W})***	&	\textcolor{white}{~}0.19 (\colorbox{lime}{W})***	&	\textcolor{white}{~}0.51 (\colorbox{yellow}{M})***\\ 
30	&	 -0.06 (\colorbox{white}{N})	&	\textcolor{white}{~}0.50 (\colorbox{yellow}{M})***	&	\textcolor{white}{~}0.31 (\colorbox{lime}{W})***	&	 -0.10 (\colorbox{white}{N})*	&	\textcolor{white}{~}0.01 (\colorbox{white}{N})	&	\textcolor{white}{~}0.37 (\colorbox{lime}{W})***\\ 
31	&	\textcolor{white}{~}0.07 (\colorbox{white}{N})	&	\textcolor{white}{~}0.37 (\colorbox{lime}{W})***	&	\textcolor{white}{~}0.15 (\colorbox{lime}{W})***	&	\textcolor{white}{~}0.12 (\colorbox{lime}{W})**	&	\textcolor{white}{~}0.17 (\colorbox{lime}{W})***	&	\textcolor{white}{~}0.29 (\colorbox{lime}{W})***\\ 
32	&	\textcolor{white}{~}0.01 (\colorbox{white}{N})	&	\textcolor{white}{~}0.54 (\colorbox{yellow}{M})***	&	\textcolor{white}{~}0.15 (\colorbox{lime}{W})***	&	\textcolor{white}{~}0.03 (\colorbox{white}{N})	&	\textcolor{white}{~}0.21 (\colorbox{lime}{W})***	&	\textcolor{white}{~}0.22 (\colorbox{lime}{W})***\\ 
\textbf{Overall}	&	\textcolor{white}{~}0.21 (\colorbox{lime}{W})***	&	\textcolor{white}{~}0.68 (\colorbox{yellow}{M})***	&	\textcolor{white}{~}0.37 (\colorbox{lime}{W})***	&	\textcolor{white}{~}0.22 (\colorbox{lime}{W})***	&	\textcolor{white}{~}0.27 (\colorbox{lime}{W})***	&	\textcolor{white}{~}0.40 (\colorbox{yellow}{M})***\\ 
\hline 
\multicolumn{7}{l}{***p \textless 0.001, **p \textless 0.01, *p \textless 0.05}\end{tabular}\end{table}

We calculate the Spearman's Correlation Coefficients of each pair of the indices given a project.
Spearman's Correlation Coefficients are selected over the Pearson's Correlation Coefficients because we find that the indices do not always follow normal distributions.
We adopt the rule of thumb in \cite{akoglu2018user,dancey2007statistics} to interpret the  correlation coefficients, i.e., the $r$ values, as summarized in Table \ref{tbl:corr_rules}.
It is worth noting that a strong correlation alone does not invalidate the design of the proposed measure because the correlations between community evolution indices are largely determined by the members' activities in real-world OSS repositories.
The phenomenon we would like to observe in this study is that whether there are occasions that the indices are weakly correlated in the real-world.
Combining the results in this study and the properties of the indices as discussed in Sec. \ref{sec:measure}, we aim at showing that the proposed indices are measuring different aspects of community evolution.

The Spearman's Correlation Coefficients together with their significance levels for all the projects are listed in Table \ref{tbl:correlation}.
We also append the results of mixing all projects' data in the last row of the table.
It can be observed that among the thirty two projects involved in our study, there are four projects, i.e., project 4, 10, 13, 31, show a weak or no correlation between all pairs of the indices, which cover both active and inactive projects (refer to Table \ref{tbl:data_set}).
And over half of the projects only show a moderate correlation between a single pair of indices.
As summarized in the last row of Table \ref{tbl:correlation}, most of the pairs show a weak, but significant correlation with $r$ values below 0.4.
The \textit{merge} and \textit{expand} indices show a moderate correlation with a $r$ value of 0.4, which is at the borderline between a moderate and weak correlation.

An interesting discovery made from the results is that the \textit{split} and \textit{merge} indices show a moderate correlation for most of projects, and even a strong correlation for five projects.
The overall correlation coefficient between \textit{split} and \textit{merge} is 0.68, suggesting a positive, moderate correlation which is statistically significant.
Recall the way we calculate community \textit{split} and \textit{merge}, and the results of concurrent validity presented above, it is clear that the two indices are capable of measuring different patterns of community evolution.
The correlation between \textit{split} and \textit{merge} is a natural result emerges from the issue and pull-request discussion activities of the community members.
A possible explanation to this positive correlation is that real-world communities often regroup themselves in OSS projects studied in this paper.
It is our future work to inspect more patterns of community evolution based on the proposed indices.

\subsection{Correlation with Team Productivity}\label{sec:productivity_result}

The last experiment demonstrates the usefulness of the proposed indices in predicting OSS project teams' productivity.
We operationalize OSS team productivity by the count of commits pushed to the repository following existing work \cite{catolino2019gender,Vasilescu2015Gender}.
We perform regression analysis with the linear mixed-effects models (LMMs) to study the correlation between the indices and team productivity.
Compared to simple linear models, LMMs allow random effects to be included in the analysis.
In this study, the random effects are the effects caused by the differences of OSS projects.
For each snapshot extracted following Sec. \ref{sec:construct_dsn_snapshot}, we obtain a sample consists of six variables as follows.
First, the outcome variable is the \textit{count of commits} in a snapshot.
Second, the aggregated indices for community \textit{split}, \textit{shrink}, \textit{merge} and \textit{expand} are used as the independent variables. 
Log transformation is performed to the above variables to stabilize variance and reduce heteroscedasticity \cite{li2020redundancy,afifi2007methods}.
Finally, the \textit{project name} is included in each sample as the indicator for random effects.

\begin{table}[!t]
	\centering
	\small
	\caption{Correlation between log transformed community evolution indices and team productivity. The later is measured by $log(\texttt{count of commits} + 0.5)$. Regression analysis with linear mixed-effects models.}
	\label{table:commit_count_model3}
	\begin{tabular}{p{0.45\linewidth}ll}
		\hline
		& \multicolumn{1}{c}{Coeffs (Errors)} & Sum Sq.      \\ \hline
		(Intercept)     & 2.9307 (0.2287)*** &            \\
		$log($\textit{split}$+0.5)$  & 1.1428 (0.04734)*** & 1565.28***  \\
		$log($\textit{shrink}$+0.5)$ & 0.2593  (0.07427)*** & 32.75*** \\
		$log($\textit{merge}$+0.5)$  & 1.0495 (0.04620)*** & 1386.35*** \\
		$log($\textit{expand}$+0.5)$ & -0.9238 (0.08525)*** & 315.46***  \\
		\hline
		\multicolumn{3}{l}{AIC=50005.521; BIC=50057.839; Accuracy=0.718; }   \\
		\multicolumn{3}{l}{$R_{m}^{2}$=0.139; $R_{c}^{2}$=0.455; Num. obs.=13018; Num. groups: \textit{project name}=32} \\ \hline
		\multicolumn{3}{p{0.9\linewidth}}{*** $p<0.001$, ** $p<0.01$, * $p<0.05$.}
	\end{tabular}
\end{table}

The results are shown in Table \ref{table:commit_count_model3}.
The variance inflation factors (VIFs) for predictors including \textit{split}, \textit{shrink}, \textit{merge}, and \textit{expand} are below 1.5, which are well below the suggested threshold of 5 in existing guidelines \cite{li2020redundancy,cohen2014applied}, suggesting that there is no strong collinearity among the predictors.
The conditional R-squared $R_{c}^{2}$=0.455 is greater than the marginal R-squared $R_{m}^{2}$=0.139, suggesting that the adoption of LMMs for regression analysis is suitable.

The results suggest, after log transformation, all the indices show a significant correlation with team productivity.
Three indices including \textit{split}, \textit{shrink}, and \textit{merge} are positively correlated with team productivity.
It is surprising to find that \textit{expand} shows a negative effect.
By further observing the coefficients and `Sum Sq.', we can find that \textit{split} and \textit{merge} are more important predictors than \textit{split} and \textit{expand}.
A possible explanation to the effects of \textit{split} and \textit{merge} is that members of communities with high degrees of \textit{split} and \textit{merge} are actively regrouping themselves to join issue and pull-request discussions which results in productive teams.
And a potential reason for the negative effects of \textit{expand} is that team performance are known to be influenced by many factors beyond team size, e.g., team diversity \cite{robert2015crowd,Vasilescu2015Gender}, developer synchronization \cite{xuan2014building}, and etc.
Growth in team size potentially increases the communication and coordination costs which reduce the working efficiency.
The above results, together with the results in Sec. \ref{sec:discriminant_validity_result},  inspire us to explore different combinations of the indices as meaningful patterns, and to conduct more studies to understand the effects of community evolution in team productivity and performance in our future work. 

In summary, the regression model achieves an accuracy (i.e., the correlation coefficient between the actual and the predicted value of the outcome \cite{r_performance}) of 0.718, suggesting that the aggregated indices of community evolution are useful predictors of team productivity.
A potential threat to the robustness of the results is that non-normality and heteroscedasticity of the residuals are detected.
We consult the literature \cite{schielzeth2020robustness,knief2021violating,gelman2006data}, and conclude that the above results are trustworthy after carefully inspecting the Q-Q diagram.

\section{Threats to Validity}\label{sec:discussion}
We present a discussion the threats to validity as follows.

\textbf{Internal Validity.}
We conduct a data-driven approach to reveal the correlation between community evolution and OSS development such as team productivity.
However, there are many other factors that are known to affect or characterize the operation of OSS teams \cite{catolino2019gender,schreiber2020social}.
As a result, the conclusions made in this paper do not fully explain how OSS teams behave or imply a causal relationship.
Instead, the proposed indices should be used as useful supplements to existing measures of DSN and community evolution.

\textbf{External Validity.}
We select thirty-two projects that cover various topics and status in our study.
However, the collection of projects is small compared to the number of projects hosted on GitHub and other platforms.
The conclusions made in this paper may not generalize well to other projects.
It is our future work to include more projects in our studies.

\textbf{Construct Validity.}
The indices are designed on top of the well-established theory of information entropy to measure community evolution with clear properties about how their values respond to parameter changes.
We evaluate the construct validity of the proposed indices in terms of concurrent and discriminant validity.
We plan to seek more criteria to validate our approach in the future.

\section{Related Work}\label{sec:related_work}
From a general perspective, this work belongs to the series of research about the social aspects of OSS development started since the beginning of the OSS movement \cite{tsay2014influence,madey2002open,hahn2008emergence}.
In this section, we introduce work closely related to this paper.

OSS projects are developed and maintained by the collaborative efforts of OSS contributors over channels such as mailing lists \cite{bird2006mining}, and on ``social coding platforms'' \cite{dabbish2012social,wessel2018power,tsay2014let} such as GitHub and GitLab.
The interactions of OSS contributors naturally generate social networks around OSS projects, which is refer to as the developer social networks (DSNs) in existing literature \cite{hong2011understanding}.
Different types of DSNs--including communication \cite{bird2006mining,weiss2006evolution,panichella2014evolution,hong2011understanding,hannemann2013community,aljemabi2018empirical,ngamkajornwiwat2008exploratory,crowston2005social,ehrlich2012all,bettenburg2011mining}, collaboration \cite{wang2007measuring,surian2010mining,joblin2017structural,hahn2008emergence,singh2010small}, and hybrid \cite{panichella2014developers,tamburri2019exploring} networks--have been extracted by mining developer mailing lists and OSS repositories to study the properties and evolution of DSNs.
The extracted DSNs are often modeled as graphs \cite{hong2011understanding,aljemabi2018empirical}.
In this work, weighted, undirected graphs are used to model DSNs that reveal the communication structures of OSS contributors, based on which a series of indices are proposed to quantify the evolution of communities inside DSNs.
It should be noted that the proposed approach can also be applied to DSNs of other types.

Technologies for social network analysis (SNA) are applied to DSNs to understand the properties, the structures, and the evolution of DSNs.
Readers can refer to \cite{schreiber2020social,herbold2021systematic,chakraborty2017metrics} for comprehensive surveys for social network and community analysis in DSNs.
Different from work that performs analysis to static DSNs \cite{bird2006mining,crowston2005social,xu2005topological,jensen2011joining,tamburri2019exploring}, this work focuses on understanding the dynamic evolution of DSNs over time.
Many studies about DSN evolution are conducted by analyzing the changes and trend in series of DSN snapshots take over time \cite{wang2007measuring,hong2011understanding,hannemann2013community,aljemabi2018empirical,mens2011analysing,huang2011analysis,joblin2017structural,ngamkajornwiwat2008exploratory}.
Network metrics, such as network diameter, shortest path between nodes, node betweenness, modularity, hierarchy, centrality, clustering coefficient, and etc, are adopted in existing work \cite{hannemann2013community,huang2011analysis,aljemabi2018empirical,joblin2017structural,pinzger2008can,chakraborty2017metrics} to assess the properties of DSNs.
There are also metrics developed exclusively to measure software communities \cite{wang2007measuring}.
Based on these metrics, statistical and data mining technologies are applied to understand the evolution of DSNs \cite{huang2011analysis,syeed2013evolution}.
This work is related to the above work in that we propose indices to quantify the evolution of DSN communities which can support further analysis including but not limited to regression analysis, and pattern detection as shown in our experiments.
Different from the above work that takes the network perspective, this work takes the community perspective \cite{aljemabi2018empirical} and analyzes the evolutionary behaviors of communities in DSNs.

The patterns adopted in this paper that describe the evolutionary behaviors of DSN communities are originally proposed in \cite{palla2007quantifying}, which involve six patterns as introduced in the rest of the paper.
These patterns are shown to be promising in describing the evolution of communities in various types of social networks such as co-authorship, phone call, and blog networks \cite{palla2007quantifying,lin2007blog}.
In \cite{hong2011understanding}, the authors adopt these patterns to characterize community evolution in OSS projects including Firefox, and Bugzilla.
Similar patterns are also adopted in \cite{aljemabi2018empirical} to describe the evolution of DSNs at the community level.
This work is closely related to the above work in that we also adopt the patterns as the basis to describe community evolution.
Different from these approaches that assign a unique, nominal pattern to a matched pair of communities in consecutive time steps, we quantify the degree of different patterns which can occur simultaneously during the evolution of the communities.
Other differences of this work from the above work include: 1) community \textit{emerge} and \textit{extinct} are defined by combinations of the indices. As a result, we only include indices for four patterns as introduced in the paper; and 2) we do not require to match communities detected in different time steps to recognize them as a single evolving community, which is a challenging task \cite{palla2007quantifying} that can lead to biases if not treated with caution.

Finally, we note that the community evolution patterns studied in this work are about the patterns introduced above, which are different from the general patterns of OSS evolution in existing work, which include studies that characterize the process of preferential attachment in OSS developers when forming DSNs \cite{weiss2006evolution}, or the stages of OSS project evolution \cite{nakakoji2002evolution}, and etc.

\section{Conclusion}\label{sec:conclusion}
In this paper, we present four entropy-based indices to measure the degree of community evolutionary behaviors with respect to patterns including \textit{split}, \textit{shrink}, \textit{merge}, and \textit{expand} to understand the evolution of DSNs around OSS projects. 
This work bridges the gap between pattern-based approaches which can provide meaningful insights about the evolution of DSNs from the community perspective, and metric-based social network analysis which can support comprehensive and quantitative analysis.
The construct validity of the proposed indices are evaluated on a real-world data set about the issue and pull-request discussions in thirty-two OSS projects.
The results suggest the indices can accurately capture different patterns of community evolution by achieving a detection accuracy of 94.1\% compared to existing work \cite{hong2011understanding}, and show weak and moderate correlations.
The results also show that the indices can accurately predict OSS team productivity after log transformation with linear mixed-effects models.

The proposed indices can be used and extended in many ways to support future studies.
For example, we can use the indices to assess and predict the behavior and status of OSS teams and projects, e.g., to assess the health of OSS teams, and to predict OSS projects' sustainability \cite{raja_defining_2012}.
More patterns of community evolution can be developed based on combinations of the indices, e.g., community \emph{regroup} as discussed in Sec. \ref{sec:discriminant_validity_result}, following the way we define community \textit{emerge} and \textit{extinct}.
We can also apply the proposed framework to other types of DSNs, e.g., collaboration networks \cite{joblin2017structural}, to understand their evolution.


\bibliographystyle{ACM-Reference-Format}
\bibliography{references}

\end{document}